# Energy Loss of a High Charge Bunched Electron Beam in Plasma

N. Barov, J.B. Rosenzweig and M.C. Thompson

*UCLA Department of Physics and Astronomy,*
*405 Hilgard Ave., Los Angeles, CA 90095-1547*

*Abstract*

There has been much interest in the blowout regime of plasma wakefield acceleration (PWFA), which features ultra-high fields and nonlinear plasma motion. Using an exact analysis, we examine here a fundamental limit of nonlinear PWFA excitation, by an infinitesimally short, relativistic electron beam. The beam energy loss in this case is shown to be linear in charge even for nonlinear plasma response, where a normalized, unitless charge exceeds unity. The physical basis for this effect is discussed, as are deviations from linear behavior observed in simulations with finite length beams.

PACS Codes: 52.40.Mj, 52.75.Di, 29.17.+w, 29.27.-a

The transfer of energy from short, intense electron beams to collective electron plasma oscillations, known as plasma wakefields, has been discussed in the context of the plasma wakefield accelerator (PWFA)[1-4]. While the original proposal for the PWFA and related concepts was in the linear regime[1,2], where the plasma oscillations can be considered small perturbations about equilibrium, highly nonlinear regimes have recently been favored[3]. For example, in the highly nonlinear "blow-out" regime[4], the plasma electrons are ejected from the channel of the intense driving electron beam, resulting in an an electron-rarefied region with excellent quality electrostatic focusing fields, as well as longitudinal electromagnetic fields, which can, in tandem, stably accelerate and contain a trailing electron beam. While the beam dynamics in this regime are linear and analytically tractable, the plasma dynamics are not, and most predictions concerning plasma behavior in the blow-out regime have been deduced from numerical simulations.

Despite the lack of analytical models for the nonlinear plasma response, it has been noted in a number of studies[5-8] that the beam energy loss rate in the PWFA blow-out regime obeys scaling usually associated with the interaction of charged particles with linear media[9]. This scaling predicts that the energy loss rate is proportional to the square of the plasma frequency[9], $\omega_p^2$. As the efficient excitation of an oscillatory system by a pulse occurs when the pulse in short compared with the oscillator period[5-8], this scaling further implies that the PWFA driving beam's energy loss rate is proportional to the inverse square of the driving beam's rms pulse length, $\sigma_z$. This prediction has led to a number of experiments that employ bunch compressors in order to decrease $\sigma_z$, thus dramatically increasing the transfer of beam energy to the plasma. In recent measurements with compressed beam at FNAL[10], the trailing portion of a 5 nC, 14 MeV, $\sigma_z$=1.2 mm, beam pulse was nearly stopped in 8 cm of $n_0$ $10^{14}$ cm$^{-3}$ plasma, a deceleration rate of over 150 MeV/m. The large collective field observed in this as well as other recent PWFA experiments[10,11], was obtained the context of nonlinear plasma electron motion, thus re-opening the issue of wakefield scaling validity. In addition, it has recently been a proposed to use ultra-short, high charge beams to drive PWFA in the tens of

GeV/m range, for creation of an ultra-high energy plasma accelerator[8,12,13]. This paper is primarily intended to address the underlying physics of the linear-like wakefield scaling of relativistic beam energy loss in plasma, and to study deviations from this scaling.

To examine the relevant physics, we perform an analysis of the motion of cold plasma electrons with equilibrium density $n_0$ (equal and opposite in charge density to the ions, which are assumed stationary) as they are excited by a relativistic ($v_b \simeq c$) electron beam. The state of plasma motion is described in terms of the velocity $\vec{v}$ and momentum $\vec{p} = \gamma m_e \vec{v}$, where the Lorentz factor $\gamma = \left[1 - (\vec{v}/c)^2\right]^{-1/2}$. The necessary relations for describing the cold plasma electrodynamic response are the Maxwell equations

$$\vec{\nabla} \times \vec{H} = \frac{4\pi}{c}\vec{J} + \frac{1}{c}\frac{\partial \vec{E}}{\partial t}, \text{ and } \vec{\nabla} \times \vec{E} = -\frac{1}{c}\frac{\partial \vec{H}}{\partial t}, \qquad [1]$$

the Lorentz force equation in convective form,

$$\frac{\partial \vec{p}}{\partial t} + \left(\vec{v} \cdot \vec{\nabla}\right)\vec{p} = -e\left[\vec{E} + \frac{1}{c}\vec{v} \times \vec{H}\right], \qquad [2]$$

and the equation of continuity for charge and current density

$$\frac{\partial n}{\partial t} + \vec{\nabla} \cdot (n\vec{v}) = 0. \qquad [3]$$

The results of our analysis will be made more transparent by the adoption of unitless variables. The natural variables used in discussing a cold plasma problem parameterize time and space in terms of the plasma frequency $\omega_p = \sqrt{4\pi e^2 n_0 / m_e}$, and wave-number $k_p = \omega_p / c$, respectively, densities in terms of $n_0$, and the amplitudes of the $\vec{E}$ and $\vec{H}$ fields in terms of the commonly termed "wave-breaking limit" $E_{WB} = m_e c \omega_p / e$. In addition, all velocities and momenta are normalized to $c$ and $m_e c$, respectively. We thus write the spatio-temporal variables, charge and current density, and field components as

$$\tilde{r} = k_p r, \ \tau = \omega_p (t - z/c), \ \tilde{v}_i = v_i / c, \ \tilde{p}_i = p_i / m_e c, \qquad [4]$$

$$\tilde{n} = n_e/n_0, \quad \tilde{J}_i = J_i/n_0 c \qquad [5]$$

$$\tilde{E}_i = E_i/E_{WB}, \quad \tilde{H}_i = H_i/E_{WB}. \qquad [6]$$

Note that use of Eqs. 4 imply that we are assuming a steady-state response (wave *ansatz*), where *t* and *z* occur only in the combination $t - z/c$. With these variables, we may write a general equation for the azimuthal component of $\tilde{H}$

$$\frac{\partial^2 \tilde{H}_\phi}{\partial \tilde{r}^2} + \frac{1}{\tilde{r}} \frac{\partial \tilde{H}_\phi}{\partial \tilde{r}} - \frac{\tilde{H}_\phi}{\tilde{r}^2} = \frac{\partial \tilde{J}_r}{\partial \tau} + \frac{\partial \tilde{J}_z}{\partial \tilde{r}}. \qquad [7]$$

In addition to the governing equation for $\tilde{H}$, we will need relationships between fields and current sources,

$$\frac{\partial \tilde{E}_z}{\partial \tilde{r}} = \tilde{J}_r \text{ and } \frac{\partial}{\partial \tau}\left(\tilde{E}_r - \tilde{H}_\phi\right) = -\tilde{J}_r. \qquad [8]$$

In this analysis the induced $\tilde{E}_z$ is found most directly by determining the transverse current, as is customary in media-stimulated radiation calculations (*cf.* Jackson, Ref. 9).

Equation 7 is nonlinear, but may be simplified by assuming small amplitude response, in which the $|\tilde{n}|$, $|\tilde{E}_i|$ and $|\tilde{H}_i|$ are small compared to unity. In fact, to place our results in perspective, we must begin with a review of previous work in the linear regime[1,2]. From the viewpoint of the fluid equations, linearity importantly implies that the plasma electron response is nonrelativistic, *i.e.* $\vec{v} = \vec{p}/m_e$. Also, in order to illustrate the dependence of $\tilde{E}_z$ on transverse beam size, we assume a disk-like beam, uniform up to radius $\tilde{a} = k_p a \ll 1$, and δ-function in $\tau$. We are interested in the instantaneous response to the plasma, and integrate over the δ-function in Eq. 8 to obtain $\tilde{H}_\phi = \tilde{E}_r$ before, during, and immediately after beam passage, which we use to further find

$$\frac{\partial^2 \tilde{H}_\phi}{\partial \tilde{r}^2} + \frac{1}{\tilde{r}} \frac{\partial \tilde{H}_\phi}{\partial \tilde{r}} - \frac{\tilde{H}_\phi}{\tilde{r}^2} - \tilde{H}_\phi = \frac{\tilde{Q}}{\pi \tilde{a}^2} \delta(\tau) \delta(\tilde{r} - \tilde{a}). \qquad [9]$$

Here we have introduced a normalized beam charge

$$\tilde{Q} = 4\pi k_p r_e N_b, \tag{10}$$

which, when $\tilde{Q} \ll 1$, indicates that the response of the system is linear. It should be noted in this regard that the experiments of Refs. 10 and 11 have beam-plasma systems yielding $\tilde{Q}$ values between 2 and 4 — values of $\tilde{Q}$ indicate that the beam is denser than the plasma as long as $k_p a$ and $k_p \sigma_z$ are below unity, as expected in the blow-out regime.

Equation 9 has a temporal δ-function which we integrate over, to obtain an inhomogenous modified Bessel equation in $\tilde{r}$

$$\frac{\partial^2 \mathbf{H}}{\partial \tilde{r}^2} + \frac{1}{\tilde{r}} \frac{\partial \mathbf{H}}{\partial \tilde{r}} - \frac{\mathbf{H}}{\tilde{r}^2} - \mathbf{H} = \frac{\tilde{Q}}{\pi \tilde{a}^2} \delta(\tilde{r} - \tilde{a}), \tag{11}$$

where $\mathbf{H} = \int_{\varepsilon-}^{\varepsilon+} \tilde{H}_\phi d\tau = \int_{\varepsilon-}^{\varepsilon+} \tilde{E}_r d\tau$. We interpret $\mathbf{H}$ as the radial momentum impulse $\tilde{p}_r$, which in the non-relativistic limit is also approximately equal to $\tilde{J}_r$ immediately behind the beam. The solution to Eq. 11 is given by

$$\mathbf{H}(\tilde{r}) = \frac{\tilde{Q}}{\pi \tilde{a}} \begin{matrix} K_1(\tilde{a}) I_1(\tilde{r}) & (\tilde{r} < \tilde{a}) \\ K_1(\tilde{r}) I_1(\tilde{a}) & (\tilde{r} > \tilde{a}), \end{matrix} \tag{12}$$

where $I_1$ and $K_1$ are modified Bessel functions.

We are interested in $\tilde{E}_z$ directly behind the beam, which is found by integrating Eq. 12

$$\tilde{E}_z(\tilde{r})\Big|_{\tau=\varepsilon+} = \int^{\tilde{r}} \mathbf{H}(\tilde{r}) d\tilde{r} = \frac{\tilde{Q}}{\pi \tilde{a}^2} \frac{1 - \tilde{a} K_1(\tilde{a}) I_0(\tilde{r})}{\tilde{a} I_1(\tilde{a}) K_0(\tilde{r})}. \tag{13}$$

For $\tilde{a} \ll 1$, the field inside of the disk is nearly constant, and given by

$$\tilde{E}_z(\tilde{r})\Big|_{\tau=\varepsilon+} \cong \frac{\tilde{Q}}{\pi \tilde{a}^2} [1 - \tilde{a} K_1(\tilde{a})] \cong \frac{\tilde{Q}}{2\pi} \left( \ln \frac{2}{\tilde{a}} - 0.577... \right), \tag{14}$$

which is to leading order proportional to $\tilde{Q}/2\pi$. In physical units we may write Eq. 14 as

$$eE_z\Big|_{\tau=\varepsilon+} \cong 2e^2 k_p^2 N_b \ln \frac{1.123}{k_p a}. \tag{15}$$

Several comments arise from inspection of Eq. 15. The first is that the scaling of $E_z$ with respect to wavenumber $k$ is dominated by the factor $k_p^2$ that is typical of Cerenkov radiation[9], if we interpret $k_p$ as the maximum allowable value of $k$ radiated. The second comment is that the linear result is ill-behaved in the limit of $k_p a << 1$, as Eq. 15 predicts a logarithmic divergence in $E_z$. This pathology is a result of allowing $\tilde{J}_r$ (through **H**) to diverge as $r^{-1}$. Previous analyses by Jackson[9] and Chen, *et al.*,[1] have attempted to mitigate this problem for the point charge limit by introducing a lower bound on $r$ (or impact parameter $b$), and have chosen the Debye length[1]. This *ad hoc* way of removing the divergence of an ultra-relativistic particle's energy loss in plasma has a dubious physical basis, however. Debye shielding places the scale of maximum distance that a particle's macroscopic field can be observed in the plasma, after a thermal equilibrium is established by the motion of the plasma electrons. However, here we are concerned with the minimum distance for which the fluid analysis is valid in describing the plasma electron response to an extremely fast transient, a particle with velocity much higher than the plasma electron thermal velocity.

As $\tilde{Q}$ is raised, we must consider the plasma electrons' relativistic response to large amplitude fields, where $\tilde{H}_\phi = \tilde{E}_r$. It is most straightforward to evaluate this response in the rest frame of the beam[14], where the beam charge gives rise to only an electrostatic field. One can then find the radial momentum kick in this frame, and Lorentz transform back to the lab fram to obtain $\tilde{p}_r$ and $\tilde{p}_z$. We therefore find

$$\tilde{p}_r = \mathbf{H}, \qquad [16]$$

while the longitudinal momentum impulse is given by

$$\tilde{p}_z = \tfrac{1}{2}\tilde{p}_r^2. \qquad [17]$$

---

[1] It is not quite clear from Jackson's discussion if the Debye length is to be applied as the lower bound to $b$ only in the case to non-relativistic particles (*cf.* Ex. 13.3, Ref. 9).

For large $\mathbf{H}$, the plasma electrons experience a large forward momentum impulse, and can have relativistic $v_z$ just after passage of the beam[2]. Equations 16 and 17 were verified by numerical integration in pulses of finite length; with their use the plasma electron's transverse velocity becomes

$$\tilde{v}_r = \frac{\mathbf{H}}{\sqrt{1 + \mathbf{H}^2 + \frac{1}{4}\mathbf{H}^4}} = \frac{\mathbf{H}}{1 + \frac{1}{2}\mathbf{H}^2}. \qquad [18]$$

In order to relate this $\tilde{v}_r$ to $\tilde{J}_r$ we must multiply by $\tilde{n}$, which due to the change in $\tilde{v}_z$ directly after passage of the beam, is predicted with the aid of Eq. 17 to be

$$\tilde{n} = (1 - \tilde{v}_z)^{-1} = 1 + \frac{1}{2}\mathbf{H}^2. \qquad [19]$$

Thus we are led to the remarkable result that the relativistically correct induced radial current is identical to the approximate, linear, non-relativistic expression,

$$\tilde{J}_r = \tilde{n}\tilde{v}_r = \left(1 + \frac{1}{2}\mathbf{H}^2\right) \frac{\mathbf{H}}{1 + \frac{1}{2}\mathbf{H}^2} = \mathbf{H}. \qquad [20]$$

Since the induced $\tilde{J}_r$ is unchanged from the linear case, the analysis of the decelerating field $\tilde{E}_z$ leading to Eq. 13 remains valid. Thus we see that the "linear" scaling observed in simulations of short pulse beam-excited wakefields may be understood partly on an analytical basis. The result in Eq. 20 arises from two effects which cancel each other: the induced $\tilde{v}_r$ saturates (at a value well below 1), yet the density enhancement due to longitudinal motion — a "snow-plowing" of the plasma electrons by the electromagnetic pressure — exactly makes up for this saturation, and the induced $\tilde{J}_r$ remains linear in $\tilde{Q}$. This snow-plowing is analogous to the scenario from laser wake-field acceleration, where the electromagnetic pressure in gradient of a short, intense laser gives rise to a density enhancement in the laser's leading edge.

---

[2] If one chooses to derive Eqs. 15 and 16 in the laboratory frame, the solution of the "impulse" equations of motion must take into account a lengthened interaction time (a finite value evaluated in the -function limit) as the plasma elections obtain longitudinal velocity, by multiplying by the factor $(1 - v_z)^{-1}$.

Our analysis has been checked with numerical integrations of the fluid equations for finite length beams, having a longitudinal charge distribution, $\rho_b(z) \propto \exp(-z^2/2\sigma_z^2)$, and taking the limit that $k_p\sigma_z \ll 1$. In order to connect with the point beam limit, and to accurately quantify the energy imparted to the plasma, we compare the average on-axis beam energy loss rate, $(2\pi\sigma_z)^{-1} \int eE_z(z)|_{r=0} \exp(-z^2/2\sigma_z^2) dz$, for these cases with linear theory. The results of these simulations are shown in Fig. 1, which displays the average energy loss of a beam in the linear regime ($\tilde{Q}$=0.002), a comparison to linear analytical theory, and the nonlinear regime ($\tilde{Q}$=2). For $k_p\sigma_z \ll 1$, the simulations converge to the linear theory, which is expected to be the case in this limit by Eqs. 16-20.

As the results of Eqs. 16-20 concern beams of negligible length they are applicable, in the limit $\tilde{a} \to 0$, to the case of a single particle. The effects of nonlinear plasma electron response do not, as might have been hoped, remove the logarithmic divergence seen in Eq. 14. Note that the logarithmic term in Eq. 15 corresponds to the familiar Coulomb logarithm[8], with an argument that is the ratio of the maximum to minimum impact parameter $b$, $\ln(b_{max}/b_{min})$. We deduce that the upper limit $b_{max} = 2/k_p$, while the lower limit in the analysis is $a$. The value of $a$ in Eqs. 13-15 cannot be drawn towards zero without violating several assumptions of our analysis, however. The fluid assumption is fine; modeling the plasma electrons as a continuous fluid introduces errors not in the average energy loss, but in the fluctuations of this quantity. For ultra-relativistic particles, quantum mechanical effects constrain the minimum impact parameter[9] to $b_{min} \simeq \frac{\hbar}{m_e c}\sqrt{\frac{2}{\gamma}}$ through the uncertainty principle, however. Thus we write the energy loss rate for a point particle of charge $q$ as

$$\frac{dU}{dz} \simeq q^2 k_p^2 \ln\left(0.794\sqrt{\gamma} \frac{m_e c}{k_p \hbar}\right) \simeq q^2 k_p^2 \ln\left(5\sqrt{\gamma} \frac{\lambda_p}{\lambda_c}\right), \qquad [22]$$

where $\lambda_p = 2\pi/k_p$ and $\lambda_c$ are the plasma and Compton wavelengths, respectively. Note that both limits of the Coulomb logarithm in Eq. 22 can be viewed quantum mechanically, as the minimum quantum of energy loss (emission of a plasmon) in the plasma is in fact $\hbar\omega_p$, as has been verified experimentally for very thin foils[15].

While the δ-function beam limit is relevant to the point-charge case, it is not of highest practical interest in bunched beams, as it has often been argued that one should set $k_p\sigma_z \approx 1$ to optimize drive beam energy loss in a PWFA[2-6,10,11]. In order to explore the deviation in plasma response from the analytical $k_p\sigma_z \ll 1$ result, we have performed a series of simulations using a fully relativistic particle-in-cell code, MAGIC[16]. Taking the beam of radius $\tilde{a} = 0.2$, and Gaussian current profile with $k_p\sigma_z = 1.1$, we have scanned the charge from linear to very nonlinear response, $\tilde{Q} = 0.02$ to 200.

The results of this parametric scan, as well as the analytical results of linear theory, are shown in Fig. 2. It can be seen that for the low amplitude cases $\tilde{Q} \ll 1$ that the linear theory predicts the average energy loss well. On the other hand, for $\tilde{Q} \gg 1$, the energy loss is significantly smaller than that predicted by linear theory, by an order of magnitude at $\tilde{Q} = 200$. Note also that the energy loss rate does not grow as $\tilde{Q}$ is increased from 60 to 200. Similar results can be deduced from other published simulations[13]. From this behavior, it can be seen that the relativistic saturation of $v_r$ and the snow-plow of the density for finite length beams do not cancel. The relative roles of these two effects, as well as that of changing plasma electron position during blow-out, can only be clarified by detailed simulation analysis. For completeness we also plot the peak accelerating field excited behind the driving beam, and its predicted value from linear theory. The peak in this field is localized in a very narrow spike, which is a small region, not terribly useful for accelerating an actual beam[8]. This spike is magnified in the more nonlinear cases, causing a field *enhancement* relative to linear theory for $\tilde{Q} \approx 1$, and partly explaining why field saturation was not noted in previous simulation scans[4,8,12]. Even with this masking effect, the accelerating peak still displays saturation when $\tilde{Q} \gg 1$.

In conclusion, we restate the most surprising of our results, that the fully relativistic response of a plasma to the passage of an ultra-short beam gives an induced electric field that is identical to the linear result. We have in the process identified a single parameter, the normalized charge $\tilde{Q}$, which identifies when a bunched beam is expected to give rise to nonlinear motion in the plasma. Further, this parameter may be used to predict the maximum physically achievable energy loss of a beam in plasma, Eq. 15. The interplay between the nonlinear effects which cancel for infinitesimal, but not for finite, length beams must be studied in more detail by simulation. Such a study is now actively under way. Even without this study, however, we may clearly state that the previously proposed scaling of wakefield amplitudes as linear with $k_p^2$ (or $\sigma_z^{-2}$ for constant $k_p \sigma_z$ and $k_p a$) is violated when the normalized charge $\tilde{Q}$ exceeds unity for finite length beams. In fact, our simulations indicate that when $\tilde{Q}$ exceeds 100, that the coupling of the beam to the plasma does not notably grow as the beam charge is increased. It should be noted in this regard that current proposals for PWFA experiments at Stanford using highly compressed beams imply beam-plasma systems with $\tilde{Q}>100$ — the prospect for observing the deviation of energy loss rate from linear scaling is therefore high.

This work supported by U.S. Dept. of Energy grant DE-FG03-92ER40693.

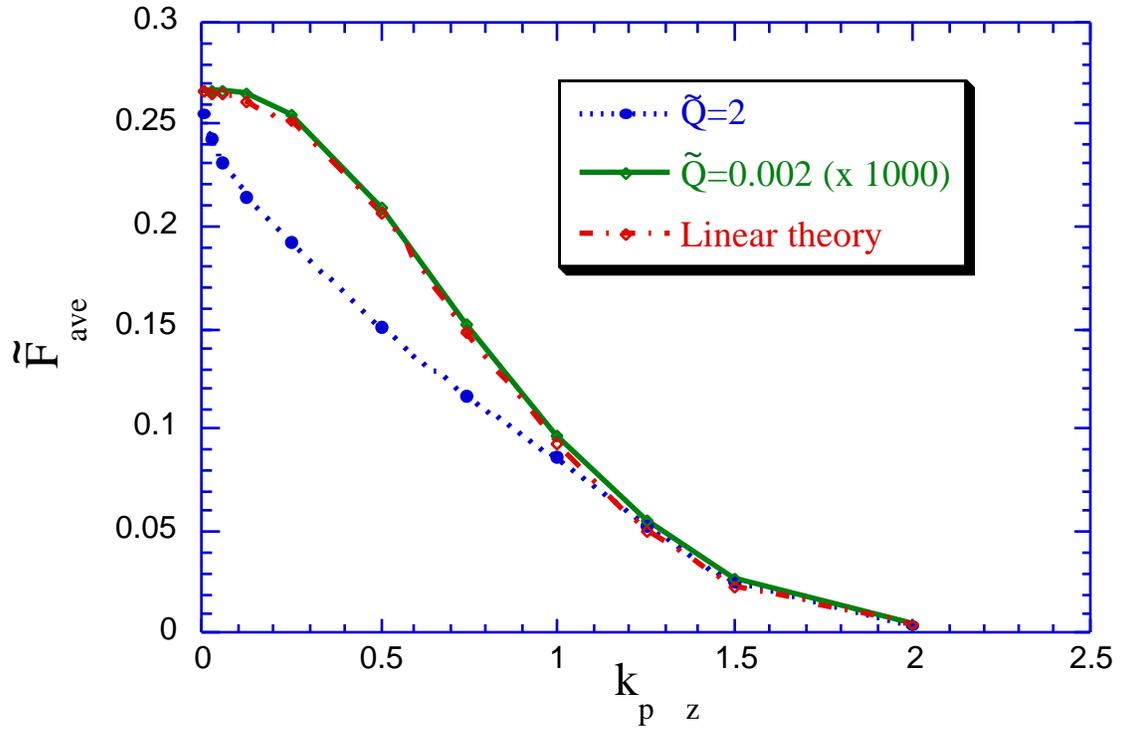

Figure 1. The average normalized energy loss rate of $\tilde{F}_{dec} = e\langle E_z \rangle / m_e c \omega_p$ of an electron beam with $k_p a = 0.2$, as a function of $k_p \sigma_z$, for $\tilde{Q}=0.002$ and 2, from cylindrically symmetric fluid simulation, and linear theory.

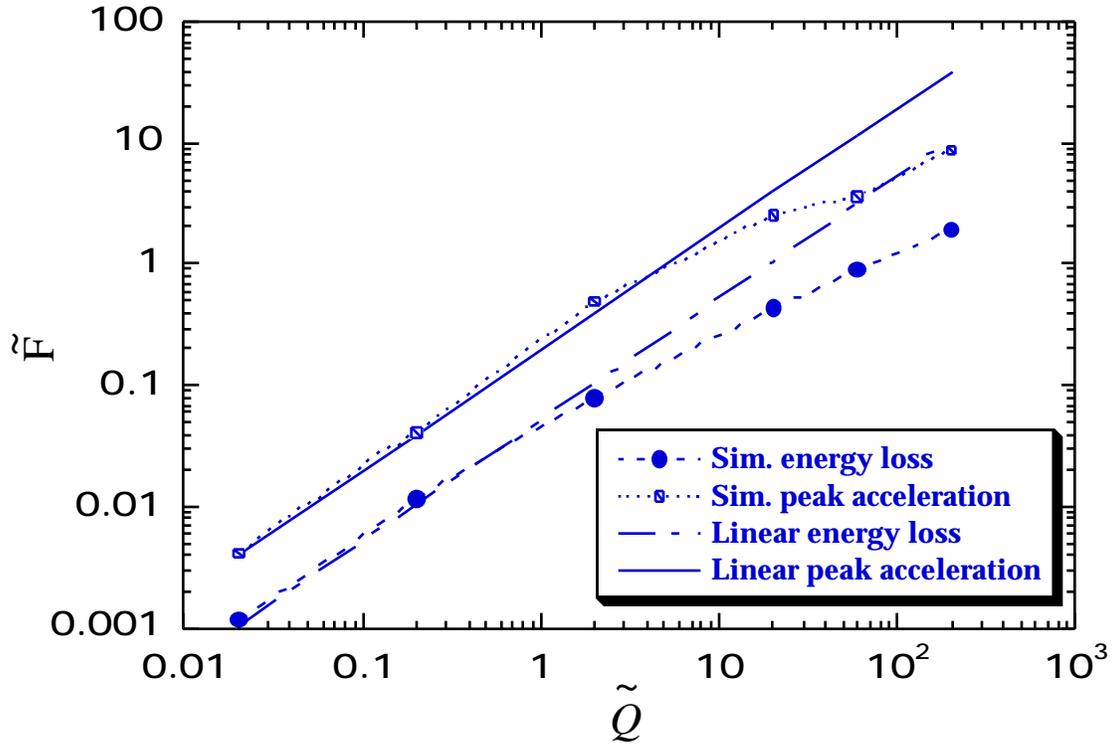

Figure 2. The average normalized energy loss rate of $\tilde{F}_{dec} = e\langle E_z\rangle/m_e c\omega_p$ of an electron beam with $k_p\sigma_z = 1.1$, $k_p a = 0.2$, as a function of $\tilde{Q}$, from linear theory and self-consistent PIC simulation; also, the peak excited accelerating field, $\tilde{F}_{max} = e|E_{z,max}|/m_e c\omega_p$, from linear theory and PIC simulation.